# The Coming Test of Social Trust in America

Elliott Middleton[1]

11 July 2017

**Abstract:** 'Animal spirits' or confidence levels are heavily dependent on how current conditions compare to adaptation levels. In the US, with its highly flexible labor markets and weak safety nets, the unemployment rate seems to serve as a generalized job insecurity indicator. In data spanning the postwar period, approximately when the unemployment rate crosses above its adaptation level as modeled by an exponential moving average over the trailing four years, a collapse of confidence occurs, and unemployment accelerates upward into a recession. In the current context of massive inequality in incomes and wealth in America, secularly declining real incomes for most Americans, and minimal trust in central government (as reflected in the election of a 'third party candidate' in the recent Presidential election) it seems likely that Americans' trust in their economic system and system of government will face severe challenges in the next cyclical downturn. If the next cycle follows previous patterns, it will be triggered by a collapse of 'animal spirits' as the unemployment rate rises above its adaptation level.

The analysis here is consistent with that of Sornette and Cauwels [1] that systems of many types creep toward instability in ways that are predictable in form if not always in exact timing. I argue, following the logic in "Adaptation Level and Animal Spirits" [2] and "Animal Spirits in America: April 2009" [3], that adaptation level effects relating to the unemployment rate drive "animal spirits" or confidence interpreted as a form of psychological wealth. Steeper discounting of losses than gains of psychological wealth attends crossing of the unemployment rate above its adaptation level. The perception of diminished wealth drives consumption and production decisions into a representative agent adaptation level theoretic criticality that causes declines in production and consumption and steep increases in the unemployment rate until people are adapted to the higher rate, and the rate peaks. It is conjectured that a similar adaptation level sensitivity and psychological wealth criticality account for much of the negative skewness of financial asset returns. The story simplifies considerably, but appears to capture important confidence dynamics well.

The original formulation of the model was normalized for the variance of the unemployment rate, but this discussion will focus simply on the difference between the unemployment rate and the adaptation level. Figure 1 presents the data for the postwar period. The positive skewedness of the unemployment rate changes is mirrored in negative skewedness of most output rate series during contractions.

---

[1] The views expressed are my own.

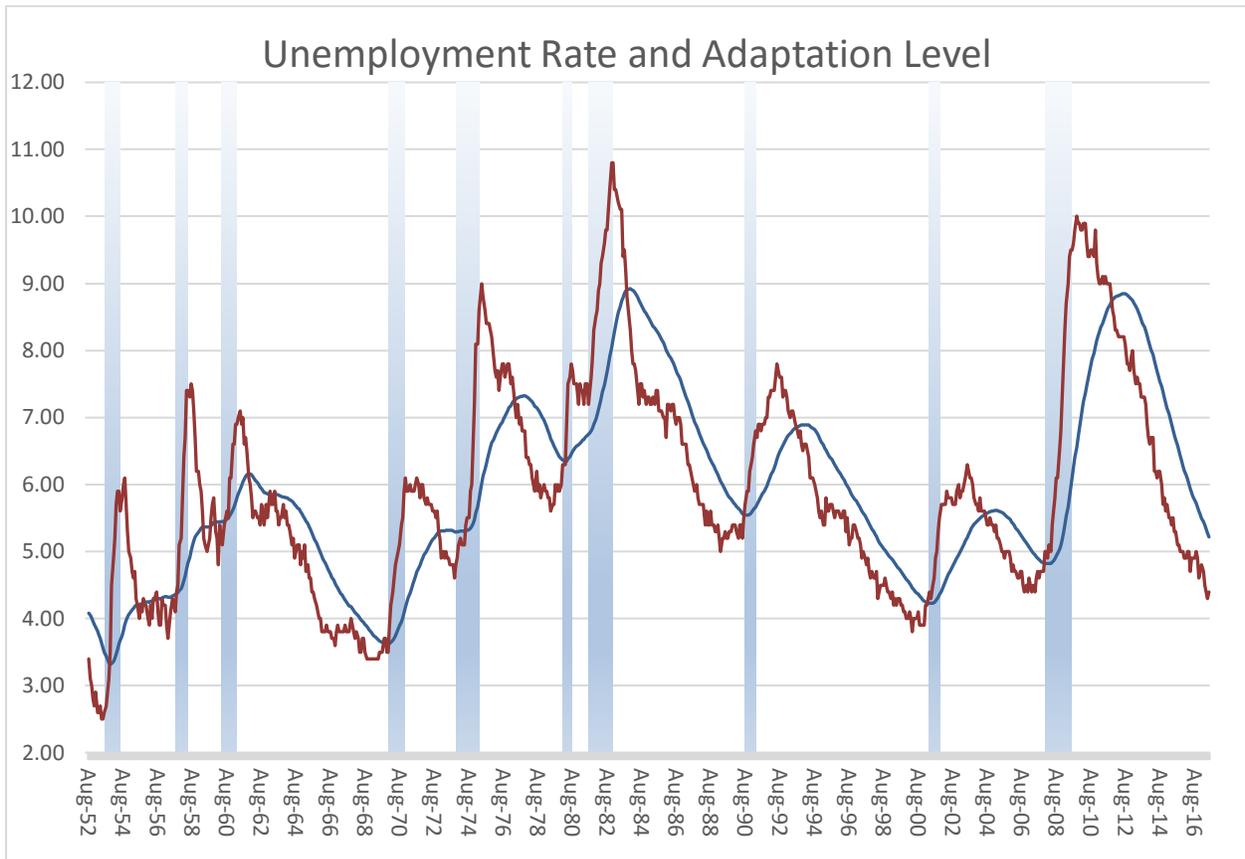

**Figure 1.** Unemployment rate and adaptation level. The onset of recessions is closely associated with the crossover point of the current rate moving above the adaptation level. See [2] for comparison with Michigan Consumer Sentiment Survey.

If a collapse of confidence is by definition a crisis, it is possible to infer that a crisis awaits at the end of every business cycle so long as there is a lower bound on the unemployment rate. It is impossible for things to get better all the time. To paraphrase Paul Volcker, the law of the business cycle has not been repealed. The Chinese seem to appreciate the essence of adaptation level theory in their philosophy of *yin* and *yang.*

Sornette and Cauwels [1] provide a number of examples of stress creep in social systems leading to disruptive instability, but the most relevant to this note is the following:

> Another remarkable case is given by Mikhail Gorbachev in a 2006 Op-ed piece for Project Syndicate […]. According to the 1990 Nobel Peace Prize winner and last head of state of the Soviet Union, the nuclear meltdown at Chernobyl, even more than the launch of *perestroika*, was the real cause of the collapse of the Soviet Union five years later. He calls the event a *Turning Point,* mentioning that *there was the era before the disaster, and there is the very different era that has followed.*

During the decades before the accident, the Soviet system had slowly drifted towards criticality, with the loss of trust functioning in a similar way as the micro-fractures in a material subject to creep. In the presence of facilitating factors, such as a weakening leadership and the accumulation of managerial mistakes, growing nationalism and exploitation by the political elites to obtain power by independence from the USSR, the weakening of the communism [sic] ideology, Glasnost's policy and economic problems, the nuclear accident contributed to tip the system over the critical threshold. The nuclear chain reaction triggered a social and political chain reaction. The catastrophe was the direct consequence of a malfunctioning organization with a flawed safety culture, poor communication and a lack of trust in its employees and it led to a general disappointment in the competence of the communist leaders, a total failure of trust in the system and contributed to the final collapse of the Soviet Union.

The first question that comes to mind when studying historical events as well as physical processes is which control parameters apply. What factors have a similar effect on social systems that temperature, pressure, or stress have on a physical system? With our explanation of the collapse of the Soviet Union, we imply that the level of trust or more generally the social capital in a society may function in that way. Support for this assumption is documented in detail by Francis Fukuyama, who described in his book *Trust* […], the impact that the *complicated and mysterious cultural process of social capital accumulation* has on different economic systems. [p. 6]

The nuclear accident at Chernobyl occurred in April 1986. The Soviet Union fell in December 1991. In Mikhail Gorbachev's mind, the accident at Chernobyl affected the control variable of social trust sufficiently to begin the descent of social trust toward collapse and delegitimization of the Soviet state. The proximate trigger for the fall of the Soviet Union in 1991 may have been radical reforms pushed by Gorbachev that caused the opportunistic secession of 11 Soviet republics to form a Commonwealth of Independent States, the three Baltic republics having already seceded [5].

The authors demolish any notion that the "Great Moderation" in America was anything more than the nation living beyond its means by easy money and excessive debt creation; see [1] and references. Picketty [4] provides evidence that institutional changes in labor and financial markets, as well as loose monetary policy and highly disparate access to credit, led to hypertrophy of financial asset returns relative to real economic growth, exacerbating inequality.

The Global Financial Crisis was America's Chernobyl. The Great Recession and the subsequent recovery comprise the weakest economic performance of the US economy since the Great Depression, "*the very different era that has followed*."

It is important to understand how the government constructs the unemployment variable. A recent study by Morningside Hill Capital Management provides details [6]:

- The Bureau of Labor Statistics (BLS) has been systemically [sic] overstating the number of jobs created, especially in the current cycle.
- The BLS has failed to account for the rise in part-time and contractual work arrangements, while all evidence points to a significant and rapid increase in the so-called contingent workforce [credible estimates are that greater than 90% of new jobs since 2008 are contingent; see [6]].
- Full-time jobs are being replaced by part-time positions, resulting in double and triple counting of jobs via the Establishment Survey.
- A full 93% of the new jobs reported since 2008 and 40% of the jobs in 2016 alone were added through the business birth and death model – a highly controversial model which is not supported by the data. On the contrary, all data on establishment births and deaths point to an ongoing decrease in entrepreneurship. […]
- Jobless claims have recently reached their lowest level in 43 years which purportedly signals job market strength. Since hiring patterns have changed significantly and increasingly more people are joining the contingent workforce, jobless claims are no longer a good leading economic indicator. Part-time and contract-based workers are most often ineligible for unemployment insurance. In the next downturn corporations will be able to cut through the contingent workforce before jobless claims show any meaningful uptick [[6], p.1].

Nevertheless, it has been the difference between the published unemployment rate and its relationship to adaptation level that has provided triggers of cyclical instability in the past, and we might expect the same to occur next time. The already insecure labor force will interpret the crossover as a signal of imminent danger to their livelihoods. Things are no longer getting better, or so they have been told; they are now definitely getting worse. The control variable of social trust may enter into a critical region if the analogy with the Soviet Union holds. Given that real median household income today is lower than in 1999 [7], and that most households' jobs and access to health care are more insecure than ever in memory, and public trust in the US central government is already at historic lows (Figure 2 [8]), the loss of social trust is likely to extend to the highly unequal form of capitalism that now characterizes the American political economy [9] as well as the central government.

Sornette and Cauwels provide a brief catalogue of systemic responses to major instability accompanying loss of social trust. The endogenous scenarios are "muddling along", "managing through" and "blood red abyss". The exogenous scenarios are "painful adjustment" and "golden east." The optimal path may be a combination of "managing through" and "golden east"; see [1] for details.

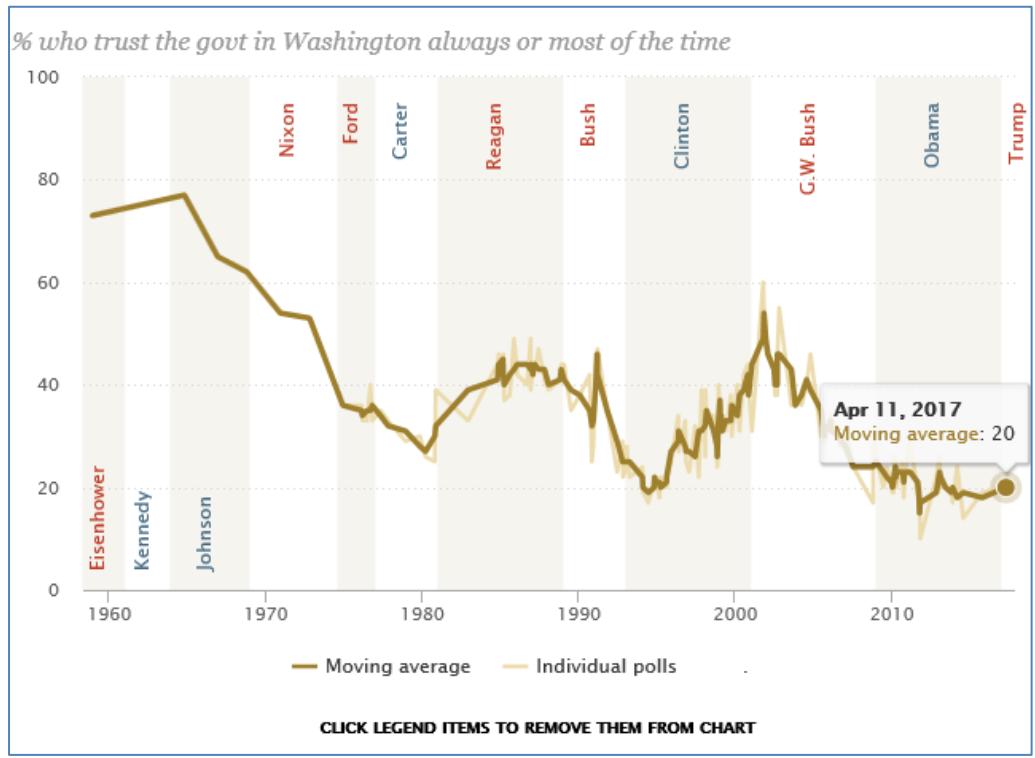

**Figure 2:** "% who trust the government in Washington always or most of the time".

**References**

[1] Sornette, D., and Cauwels, P., A Creepy World, arXiv:1401.3281v1 [physics.soc-ph], 14 Jan 2014, retrieved at https://arxiv.org/abs/1401.3281 .

[2] Middleton, E., Adaptation Level and Animal Spirits, *Journal of Economic Psychology,* **17,** 479-498, 1996.

[3] Middleton, E., Animal Spirits in America, April 2009, arXiv:0904.1431 [nlin.AO], 8 Apr 2009, retrieved at https://arxiv.org/abs/0904.1431 .

[4] Picketty, T., Capitalism in the Twenty-First Century, Harvard University Press, 2014.

[5] The History Channel, http://www.history.com/topics/cold-war/fall-of-soviet-union .